\documentclass{article}

\usepackage[utf8]{inputenc} 
\usepackage[T1]{fontenc}    
\usepackage{hyperref}       
\usepackage{url}            
\usepackage{booktabs}       
\usepackage{amsfonts}       
\usepackage{nicefrac}       
\usepackage{microtype}      
\usepackage{xcolor}         

\usepackage{longtable}
\usepackage{multirow}
\usepackage{amssymb}
\usepackage{color}
\usepackage{graphicx}
\usepackage{booktabs}
\usepackage{algorithm}
\usepackage[noend]{algpseudocode}
\usepackage{caption}
\usepackage{rotating}
\usepackage{paralist}
\usepackage{setspace}
\usepackage{amsmath} 
\usepackage{amsthm}
\usepackage{enumitem}  
\usepackage{tablefootnote}
\usepackage{dsfont}

\usepackage{pifont}

\renewcommand{\epsilon}{\varepsilon}

\newcommand{\ContrCapAlg}{\texttt{DistributedContributionBounding}}

\begin{document}

\title{Scalable contribution bounding to achieve privacy}

\author{Vincent Cohen-Addad, Alessandro Epasto, Jason Lee
 \and Morteza Zadimoghaddam \\ 
 Google Research\\ \texttt{\{cohenaddad,aepasto,jdlee,zadim\}@google.com}}

\maketitle

\begin{abstract}
In modern datasets, where single records can have multiple owners, enforcing user-level differential privacy requires capping each user's total contribution. This "contribution bounding" becomes a significant combinatorial challenge. Existing sequential algorithms for this task are computationally intensive and do not scale to the massive datasets prevalent today. To address this scalability bottleneck, we propose a novel and efficient distributed algorithm. Our approach models the complex ownership structure as a hypergraph, where users are vertices and records are hyperedges. The algorithm proceeds in rounds, allowing users to propose records in parallel. A record is added to the final dataset only if all its owners unanimously agree, thereby ensuring that no user's predefined contribution limit is violated. This method aims to maximize the size of the resulting dataset for high utility while providing a practical, scalable solution for implementing user-level privacy in large, real-world systems.
\end{abstract}

\section{Introduction}\label{Introduction}
An effective method for safeguarding individual privacy in algorithmic systems is the principle of differential privacy (DP), which mathematically guarantees that the output of an algorithm is not overly sensitive to the data of any single individual. A foundational step in many differentially private systems, particularly in complex data ownership scenarios, is to explicitly limit, or "cap," the contribution of each user. This is achieved by selecting a subset of the original data such that no user is associated with more than a predefined number, b, of records in this subset. This process, known as user contribution bounding, is crucial for mitigating privacy risks like memorization and provides a direct lever to balance the trade-off between data utility and privacy loss, for instance in event-level DP training pipelines.

In settings with \textit{single-owner data}, where each record is owned by exactly one user, this bounding process is straightforward. As users do not share records, each can independently select up to b of their own records to contribute, without impacting the selections of others \cite{charles2024finetuninglargelanguagemodels}. However, modern datasets are often characterized by complex, overlapping ownership structures where a single data record—such as a photograph with multiple people or a shared document—can be associated with several users. This multi-owner setup fundamentally complicates the contribution bounding problem, turning it into a challenging combinatorial optimization task.

Previous work has addressed this challenge in the multi-owner context. Notably, \cite{ganesh2025itsdatatooprivate} introduced a sequential Greedy algorithm to solve this problem. While they demonstrated that their approach yields results close to the optimal solution achievable via linear programming, its sequential nature imposes significant computational burdens. For the massive datasets prevalent today, sequential algorithms suffer from prohibitive run times, creating a critical need for more scalable solutions.

We present a scalable, distributed algorithm designed to address the user contribution bounding problem. We model the intricate ownership structure of the data as a hypergraph, where users are represented as vertices and data records as hyperedges connecting their respective owners. 
Our primary objective is to select the largest possible subset of these hyperedges (records) while strictly adhering to the constraint that each vertex (user) is part of no more than b selected hyperedges. 
We formalize this as the following optimization problem: given a hypergraph $G=(V,H)$ and a capacity $b$ for each user $u \in V$, find a subgraph $G =(V,M)$ with $M \subseteq H$ 
that maximizes $|M|$  subject to the constraint that the degree of 
every user $u \in G$ 
is at most $b(u)$. While we assume a uniform capacity $b$ for simplicity, our algorithm readily generalizes to non-uniform, user-specific capacities.

To solve this problem efficiently on large-scale datasets, we design a distributed algorithm, \ContrCapAlg{} (Algorithm~\ref{alg:contribution-bounding}), that operates in rounds. The algorithm allows users to collaboratively select a valid subset of records in parallel, drastically reducing the required computation time. In each round, users who have not yet reached their contribution capacity propose their most preferred records. A record is definitively selected for the final dataset only if all of its owners unanimously propose it in the same round. This approach ensures that the resulting dataset respects all user contribution bounds while enabling efficient, decentralized execution. Our work provides a practical and scalable solution for a fundamental problem in the deployment of user-level differential privacy for complex, modern datasets.

\section{Distributed User Contribution Bounding Algorithm}\label{user-contrib-bounding}
A crucial step for ensuring algorithmic privacy is to cap the output's sensitivity to any individual's data. To do this, we limit the overall contribution of each user by selecting a smaller, representative dataset from the original input. This selection process ensures that no single user is associated with more than $b$ records in the chosen data.

 We propose Algorithm~\ref{alg:contribution-bounding} to solve this contribution bounding optimization problem in a distributed manner. We model the ownership structure with a hypergraph between the users with each record represented by a hyperedge consisting of its participant users. 

Our goal is to select a subset of the input records/hyperedges in order to bound the user contribution of each user to the selected hyperedges. For utility purposes, we want to retain as much data as possible. The records selected are the output of the system. Here $b$ is the participation budget that can be fine tuned to mitigate the risk of memorization and balance the utility / privacy loss tradeoff for instance in event-level DP training pipelines. In a more general setting, we may also allow each user $u$ to have a separate capacity $b(u)$. For simplicity we will assume that $b(u) = b$ for all nodes. Our algorithm generalizes to the case of non-uniform capacities for users. We formalize the the optimization problem as follows. 

\begin{algorithm}
\caption{\ContrCapAlg}\label{alg:contribution-bounding}
\begin{algorithmic}[1]
\Require Hyper Graph $G=(V,H)$, capacities $b(u)$ and preference ordering $<^u$ over the edges for every $u \in V$, and number of rounds $R$
\Ensure Subset of hyper edges $M \subset H$.
\State $S(u) \gets \mbox{Unsaturated}$, $\forall u \in V$ 
\State $d(u) \gets 0$, $\forall u \in V$ 
\State $M \gets \emptyset$
\State $E \gets H$
\For{$r = 1$ to $R$}
  \For{user $u \in V$ with $S(u)  = \mbox{unsaturated}$}
    \State Propose $b(u) - d(u)$ top eligible (in $E$) neighbor edges of $u$ based on ordering $<^u$ for the matching.  
  \EndFor
  \For{every $h \in H$ that has been proposed by all its users}
    \State Add $h$ to $M$.
    \State Remove $h$ from $E$.
    \State Update all $d(u)$ and $S(u)$ variables for each $u$ in hyperedge $h$.
  \EndFor
  \For{For any new saturated user $u$}
    \State Remove all edges of $u$ from $E$.
  \EndFor
\EndFor
\State \Return $M$
\end{algorithmic}
\end{algorithm}

The input consists of: 
\begin{itemize}
    \item Hyper Graph $G=(V,H)$ with vertex set $V$ and set of hyper edges $H$. Each edge $h \in H$ is a subset of vertices $V$ potentially with parallel hyper-edges meaning hyper-edges that have the same subset of vertices. 
    \item Capacity $b(u)$ with the number of hyperedges allowed for every $u \in V$. 
\end{itemize}

For the output, we expect a subset $M \subset H$ of the hyper-edges.
The objective is
maximizing the cardinality of $M$.
As the feasibility constraint, we require that in
the sub-hypergraph $G' = (V, M)$ no user $u$ has degree larger than $b(u)$.  

Since the algorithm needs to scale to massive datasets, we focus on designing distributed algorithms that can finish in a few rounds.

Our \ContrCapAlg{} formalized as Algorithm~\ref{alg:contribution-bounding} proceeds in rounds. In each round the nodes have a state of $S(u)$ indicating if the node is Saturated, or Unsaturated.  All nodes start in the state of Unsaturated.
Each node $u$ also keeps the count $d(u)$ of the matched hyperedges for $u$ so far. This count is initialized with zero.

The algorithm continuously updates a b-matching $M$ which is a subset of edges respecting the capacity constraints. $M$ is initialized with an empty set and is guaranteed to respect the constraint at all times.  
 
Our algorithm further keeps a subset of hyperedges $E \subset H$ as eligible edges. $E$ is initialized with all hyperedges $E = H$.

We also assume some user-specific ordering ($<^u$) over the hyper edges. This can be using their weights (if such weights are provided) or it can be optimized in any way. The default behavior is to generate a universal random ordering of hyperedges that is shared by all users, and hence is consistent across all users. Such ordering can be generated via a random hash function that is seeded with a unique edge id. 

In each round, all vertices with $S(u) = \mbox{unsaturated}$, select $b(u) - d(u)$ top neighbor hyperedges from $E$, and propose these edges for the match. Here the notion of top neighbors is defined based on  the preference order $<^u$.
A proposed hyperedge is added to $M$ if and only if all neighbors/users of the edge unanimously propose it. If the edge is added to $M$, it is removed from $E$.
All nodes update their $d(.)$ value and state $S(.)$ with the new matches. Edges that are adjacent to saturated vertices are removed from $E$. 

The algorithm outputs the union of all edges in $M$ obtained in the various rounds and can be iterated an arbitrary number of times.

\bibliographystyle{alpha}
\bibliography{references}

\end{document}